\documentclass[11pt]{article}

\usepackage{physics}
\usepackage{graphicx}
\usepackage{color}

\begin{document}


\title{A perspective on Lindblad's \textit{Non-Equilibrium Entropy}}

\author{
Erik Aurell \\{\footnotesize\it  Department of Computational Science and
  Technology, AlbaNova University Center}\\
{\footnotesize\it SE-106 91 Stockholm, Sweden \& eaurell@kth.se}\\[2ex]
Ryoichi Kawai \\{\footnotesize\it Department of Physics, University of
  Alabama at Birmingham} \\
{\footnotesize\it Birmingham AL, 35294, USA \& kawai@uab.edu} }

\maketitle

\abstract{
G\"oran Lindblad in 1983 published a monograph on non-equilibrium thermodynamics. 
We here summarize the contents
of this book, and provide a perspective on its relation to later developments in
statistical physics and quantum physics. We high-light two aspects. The first is
the idea that while all unitaries can be allowed in principle, different theories result
from limiting which unitary evolutions are realized in the real world.
The second is that Lindblad's proposal for thermodynamic entropy
(as opposed to information-theoretic entropy) foreshadows much more recent investigations
into optimal quantum transport which is
a current research focus in several fields.
\\\\
This paper is an Invited contribution to the Lindblad memorial volume, to be published in Open Systems and Information Dynamics (2023).
}

\section{Introduction}
\label{sec:intro}

The highest impact work of G\"{o}ran Lindblad is his work on quantum dynamical semigroup, published in 1976~\cite{Lindblad1976}. Scopus reports that the paper was cited more than 4800 times and Google Scholar indicates more than 7700 times. So-called Lindblad equation and Lindblad generators are extensively used today, nearly a half a century after the publication. Without a doubt, Lindblad was a forerunner of research on open quantum systems, as he was also in quantum information theory.

\begin{figure}
    \centering
    \includegraphics[width=3in]{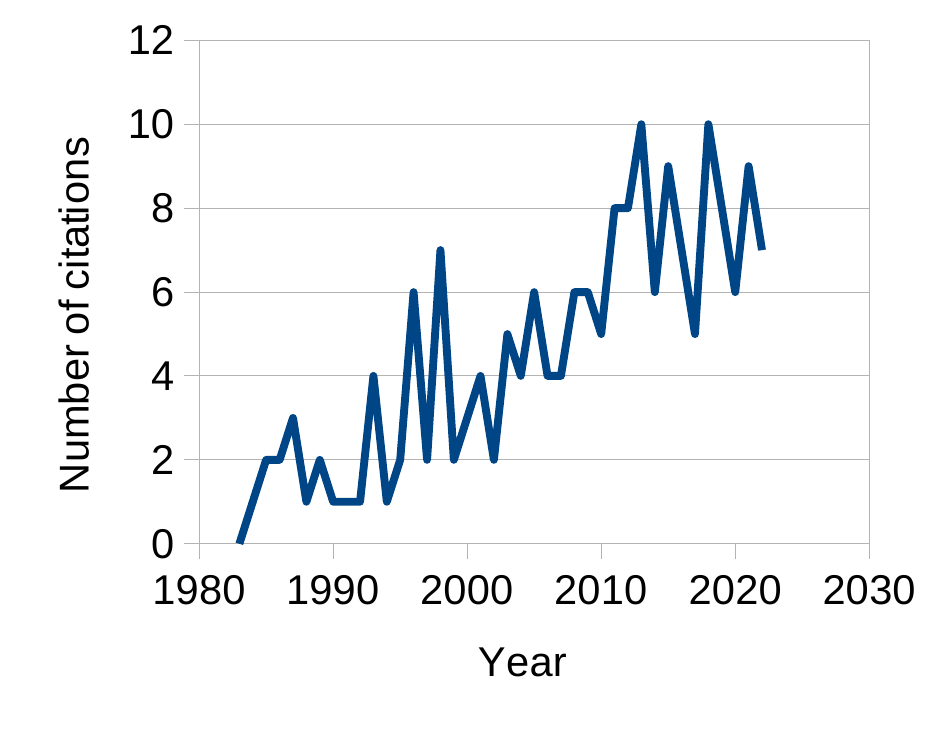}
    \caption{The number of citations of Lindblad's monograph~\cite{Lindblad1983} per year.  The citation count includes research articles in journals, proceedings, and arXiv, books, and Ph.D dissertations. }\label{fig:citation}
\end{figure}

A decade later
Lindblad in 1983 published a monograph "Non-Equilibrium Entropy and Irreversibility"~\cite{Lindblad1983}. The book has to date been cited about 200 times, predominantly though not exclusively in the quantum thermodynamics literature. Figure \ref{fig:citation} shows that the book is more often cited in the 21st century, indicating that this work was also ahead of its time.
This contribution to the memorial volume is about Lindblad's monograph from the perspective of modern statistical physics. 
By providing an extensive synopsis we will show that Lindblad already then
was pursuing the idea that while all unitaries can be allowed in 
quantum theory in principle, different theories result
from limiting which evolutions are realized in the real world.
The same principle underlies Lindblad's last paper \cite{Lindblad2022} (this volume).

As the title suggests, the book is about the origin of irreversibility and the thermodynamic entropy of non-equilibrium states. While many ideas introduced in the book can be applied to both classical and quantum systems, Lindblad uses the language of quantum mechanics throughout the book and the underlying quantum evolution is mainly based on the quantum dynamical semigroup he had developed. He tells us that elusive concepts, namely, irreversibility, thermalisation, and the second law of thermodynamics, can be explained by quantum Markov processes.  In one sense, the book can be seen as an application of his theory of open quantum systems to quantum thermodynamics.  However, it is far beyond that. The main goal of the book is to construct a thermodynamic entropy function for non-equilibrium states.  He carefully develops necessary concepts and tools, step by step.  We can recognize many pioneering ideas (and also struggles) in quantum thermodynamics which became a major research field of physics in the 21st century.
More specifically, Lindblad's proposal for non-equilibrium thermodynamic entropy
(as opposed to information-theoretic entropy) is related to the problem of optimal quantum transport under constraints, a key research question in several areas of open quantum systems theory, and with many applications to current and future quantum technologies.

\section{A synopsis of the book}\label{sec:synposis}

In the following subsections we give an extended synposis of each chapter in the book, providing a background of the discussion, interpretation of his ideas, a brief summary of his conclusions, and comparison between the Lindblad's work and later developments.

\subsection{Preface and Chapter 1 "Introduction and Summary"}\label{sec:synopsis-1}

Lindblad begins the preface by stating that the problem of deriving irreversible thermodynamics from reversible microscopic dynamics has been on the agenda of theoretical physics for more than a century. He states his ambition to present a different approach of which the key aspects are
\begin{itemize}
    \item that the relation between dynamics and thermodynamics can be based on the concepts of energy and work, such that entropy is related to available work;
    \item that the second law of thermodynamics, if it is to have a universal validity, must be a tautology in a certain sense.
\end{itemize}
In Lindblad's own words the essential idea is expressed as
\begin{quote}
"The entropy function is not unique. Instead there is a family of such functions, one for each set of thermodynamic processes allowed by the experimenter’s control of the dynamics of the system through the external fields." (page 5)
\end{quote}

It is striking that a scientist who \textit{inter alia} is famous for deriving entropy inequalities at the birth of quantum information theory (see the profile of the life and career of Lindblad by Ingemar Bengtsson \cite{Bengtsson-profile}, elsewhere in this issue) emphatically excludes an information-theoretical interpretation of entropy
for non-equilibrium states. It is also striking that in the modern theory of stochastic thermodynamics the Second Law indeed appears as a consequence of an equality (a Fluctuation Relation),
which in many systems holds as a tautology \cite{Jarzynski2011,Kawai2007,Parrondo2009,Sevick-Searles2008,Seifert2012}.
The reader in his or her mind may thus form the hypothesis that perhaps Lindblad was an unrecognized precursor of much later developments, which have only recently risen to
dominance in statistical physics. This review has been written with such an hypothesis in mind.

Lindblad defines non-equilibrium entropy in a manner which is partly modern, and partly restricted to specific models
\begin{quote}
"The entropy of a non-equilibrium state is defined as the infimum of the entropies of the equilibrium states which can be reached from the given state using the semigroup of evolutions generated by the Hamiltonian dynamics and the interaction with heat baths." (page 5)
\end{quote}

The previous quote on multiplicity of entropy functions refer to different sets of Hamiltonians that can act on the system, corresponding to different sets of external controls.
Later in Chapter 10, he comments on this multiplicity as "At first sight, this seems disturbing in view of the unicity of the entropy of classical thermodynamics.  However, ...."  From the whole of the monograph it is clear that non-uniqueness is a key ingredient in Lindblad's theory of non-equilibrium entropy. We will now, chapter by chapter, outline how Lindblad works out these concepts.
\footnote{We shall use the same mathematical notations as appeared in the book unless otherwise is noted.}

\subsection{Chapter 2 "Dynamics and Work"}\label{sec:synopsis-2}

In Chapter 2 of the book, the dynamics controlled by an experimenter is mathematically defined as the first step towards the multiplicity of entropy in his theory.
Let $F$ be a set of Hamiltonian operators with external fields the experimenter controls, which act on the system.   The experimenter can pick multiple Hamiltonians from $F$ and construct a time-dependent Hamiltonian (piece-wise constant) by switching from one Hamiltonian to another, one by one. In particular, Lindblad defines time-dependent Hamiltonians for a duration $[s,t]$ such that Hamiltonian comes back to the origin, i.e., $H(s)=H(t)=H$.  Such a protocol is called a \emph{work cycle} and denoted as $\gamma$, where $H$ is called the origin of the work cycle. The experimenter can pick a work cycle from the set of all possible work cycles $\Gamma(F)$.  If the experimenter has a smaller number of options ($F' \subseteq F$), then obviously the number of possible work cycles is also reduced ($\Gamma(F') \subseteq \Gamma(F)$). The set of accessible states hence depends on the set of controls $F$. This \emph{F-accessibilty} is the key ingredient of the Lindblad's theory of irreversibility, and appears throughout the book.

Once the experimenter has chosen a work cycle $\gamma$, a dynamical semigroup $T(\gamma)$ (either unitary or non-unitary) maps an initial state $\rho(s)$ to a final state $\rho(t)$. The set of all dynamical semigroups allowed by $F$ is called the \emph{mobility semigroup} and denoted as $T(F)$. In fact, if $F$ contains all self-adjoint operators and all the evolutions are unitary,  then $T(F)$ contains both $U$ and $U^{\dagger}$, hence a reversible microscopic system is realized.

Furthermore, Lindblad introduces $\Omega(\rho,F)$, the set of all final states accessible from an initial state $\rho$ by $T(F)$ which satisfies crucial properties\footnote{Lindblad here orders $F$ and $F'$ oppositely. To preserve parallelism with the preceding argument and increase readability we have switched the order.}:
\begin{equation}\label{eq:subset-Omega-F}
    \Omega(\rho,F') \subseteq \Omega(\rho,F)
\end{equation}
for $F' \subseteq F$
and
\begin{equation}\label{eq:subset-Omega-mu}
    \Omega(\mu,F) \subseteq \Omega(\rho,F)
\end{equation}
for $\forall \mu \in \Omega(\rho,F)$. These relations are used later in the book to show irreversibility.

The \emph{work} performed during a work cycle $\gamma$ is defined as\footnote{Throughout the book Lindblad uses the notation $\rho(X) \equiv \Tr[\rho X]$ where $X$ is an operator.  For readability, we shall write the trace explicitly.}
\begin{equation}\label{eq:work-definition}
  W(\gamma)= \int_\gamma  \rho(u) [dH(u)] = \int_s^t \Tr\left[\rho(u) \pdv{H(u)}{u}\right] du.
\end{equation}
While this definition is general, it is not quite practical since the time evolution of the state must be known and the trace must be integrated.
For unitary evolutions  we can do an integration by parts and use the von-Neumann-Liouville equation for $\dot{\rho}$ to obtain
\begin{equation}
  \label{eq:work-two-point}
  W(\gamma,\rho) = \Tr[\rho \hat{W}(\gamma)]
\end{equation}
where $\hat{W}(\gamma) = \hat{H}(s) - \hat{H}(t)$ with the Hamiltonian $\hat{H}$ expressed in the Heisenberg picture. This definition is now known as the two-point measurement formalism.

Lindblad appears to have taken for granted that $\hat{W}$ defined by Eq.~\eqref{eq:work-two-point} is a quantum-mechanical observable. In quantum thermodynamics this has been questioned since $\hat{W}$ is not a standard observable~\cite{Talkner2007}\footnote{There is a recent report that work can be a quantum operator and its eigenvalues have physical meaning \cite{PintoSilva2021}.
It is also possible to carry forward a first measurement by the use of ancillas so that the generating function of work defined by the two-measurement
scheme is measurable~\cite{Mazzola2013}, a point subsequently made more generally by considering positive operator valued measures (POVM)~\cite{Roncaglia2014}.
}. Apart from the 
perhaps mis-leading word choice, the definition itself is however reasonable, and widely used to date. 

For non-unitary evolutions the situation is more complex~\cite{PerarnauLlobet2017,Talkner2009}. If the evolution is unital, Eq.~\eqref{eq:work-two-point} 
can still be used~\cite{Albash2013,Rastegin2014}. For general non-unitary evolution, one approach is just to integrate Eq.~\eqref{eq:work-definition} with the quantum state $\rho$
obtained from a non-unitary evolution, e.g., a Lindblad equation. This does not lead to Eq.~\eqref{eq:work-two-point}, and we must carry out the integral for each specific time-dependent control Hamiltonian $H$.
We can also represent non-unitary evolution as unitary evolution of the system and an appropriately chosen environment such that in Eq.~\eqref{eq:work-definition}
$\rho$ is the total density matrix. Then Eq.~\eqref{eq:work-two-point} holds for the total Hamiltonian of the system and the environment
and work so defined includes both change of internal energy and change of environment energy. The latter can be assimilated to heat and recalls the First Law (in expectation). Unfortunately, this definition supposes that expected energy changes in the environment are measurable, which is questionable from a strict thermodynamic point of view, and
often difficult to execute in practice.

There are still more open questions with the definition of work for quantum thermodynamics.  Measuring energy is not necessarily free of cost. When the measurement is destructive, we must take into account quantum back action.  Furthermore, the exchange of energy and information between the system and the measurement device needs to be included~\cite{Deffner2016,Hanggi2015,Sone2020,Talkner2016}.  Lindblad discusses the effects of quantum measurement on work extraction in Chapter 11.

We end this section by emphasizing again that Lindblad throughout assumes that a work cycle $\gamma$ is constructed from a set $F$. The mobility semigroup $T(F)$ hence consists of the F-accessible Lindblad-equation evolutions, and $\Omega(\rho,F)$ is the set of states that can be reached from an initial state $\rho$ evolving under an element in $T(F)$.

\subsection{Chapter 3 "Information entropy"}\label{sec:synopsis-3}

In this chapter Lindblad introduces the basic concepts of (quantum) information theory, a set of tools, concepts and results now widely known~\cite{Nielsen2000,Wilde2013}. First, he introduces two entropies, von Neumann entropy  $S_I(\rho)=-\Tr (\rho\ln\rho)$ (Lindblad calls it I-entropy) and relative entropy $S_I(\rho|\mu) = \Tr(\rho\ln\rho) - \Tr(\rho\ln\mu)$.  Then, he provides a summary of their basic properties, namely, positivity, unitary invariance, concavity, and subadditivity for the von Neumann entropy and positivity, unitary invariance, and  data processing inequality (Lindblad calls it quantum H-theorem) for the relative entropy.

From the perspective of non-equilibrium thermodynamics, the unitary invariance disqualifies the von Neumann entropy as a candidate of thermodynamic entropy since it cannot describe irreversibility.  The quantum H-theorem has thermodynamic significance an the understanding of irreversibility, which Lindblad discusses in Chapter 10.

The information entropies are more useful for equilibrium states.  First, 
Lindblad shows that thermodynamic entropy $S[\beta,H]$ of equilibrium states $\rho(\beta,H)$, coincides with the information entropy of the Gibbs state $S_I(\rho(\beta, H))$.  Then, he introduces a useful equality involving a Gibbs state. For any state $\rho$,
\begin{equation}\label{eq:thermo_stability}
    S_I[\rho|\rho(\beta,H)] = S(\beta,H) - S_I(\rho) + \beta[\textrm{Tr}(H \rho) - \textrm{Tr}(H \rho(\beta,H))] \ge 0 ,
\end{equation}
which leads to the variational principles of thermodynamics:
\begin{subequations}
\begin{equation}
    S(\beta,H) = S_I(\rho)  \quad \Rightarrow \quad \Tr(\rho H) \ge \Tr[\rho(\beta,H)]
\end{equation}
\begin{equation}
    \Tr(\rho H) = \Tr[\rho(\beta,H)]  \quad \Rightarrow \quad  S(\beta,H) \ge	S_I(\rho).
    \end{equation}
\end{subequations}
Equation \eqref{eq:thermo_stability} is particularly useful when 
Lindblad considers the thermodynamic limit in Sec.~\ref{sec:synopsis-5}.

For unitarily evolving states $\rho(t)$, the work as defined by Eqs.~\eqref{eq:work-definition} and \eqref{eq:work-two-point} can be expressed 
using relative entropy as
\begin{equation}\label{eq:work-S}
    W\left(\gamma,\rho(s)\right)=\beta^{-1}\left(S_I\left[\rho(s) | \rho(\beta,H)\right]
        - S_I\left[\rho(t) | \rho(\beta,H)\right]\right)
\end{equation}
where $\gamma$ is the work cycle for period $[s,t]$. From equality \eqref{eq:work-S} and the properties of relative entropy, Lindblad draws the following three conclusions:
\begin{itemize}
    \item [(1)] If $\rho(s)$ is an equilibrium state then the work is non-positive, consistent with the passivity of the Gibbs state.
    \item [(2)] If $W=0$ then $\rho(t)$ is the same equilibrium state as $\rho(s)$.
    \item [(3)] Since $S_I\left[\rho(t) | \rho(\beta,H)\right] \ge 0$, an upper bound of the work is given by \\$W \le \beta^{-1} S_I\left[\rho(s) | \rho(\beta,H)\right]$ and the equality holds if and only if $\rho(t)=\rho(\beta,H)$.  It follows that if an equilibrium state $\rho(\beta,H)$ can be reached by unitary evolution from some initial state $\rho(s)=\rho$, then the maximal work which can be extracted from $\rho$ in such a work cycle $\gamma$ is given by
    \begin{equation}
        \label{eq:available-work-A}
        A(\rho;H)=\beta^{-1} S_I\left[\rho | \rho(\beta,H)\right]
    \end{equation}
\end{itemize}
and that the maximum work can be obtained if and only if the final state is the Gibbs state $\rho(\beta,H)$.

As Lindblad remarks, this property had a few years earlier led Procaccia and Levine~\cite{Procaccia1976} to propose $A(\rho;H)$ as a definition of available work in an arbitrary
non-equilibrium state, for classical systems.  It was recently used as the upper bound of work in the context of quantum thermodynamics~\cite{Aberg2013,Aberg2014,Hovhannisyan2020}.  If justified this would make work practically a state functional, and so would be a highly non-trivial (and very useful) property. However, as Lindblad then goes on to remark, in general no equilibrium state is reachable by only unitary operations. In the final subsection of this chapter
Lindblad discusses entropy-increasing processes, as a preparation to the ideas developed in the later chapters.

\subsection{Chapter 4 "Heat baths"}
\label{sec:synopsis-4}

To construct a thermodynamic entropy function for non-equilibrium states, we must clearly define a heat bath or otherwise we may end up with a tautology.  Lindblad sets two required properties for idealized heat baths:
\begin{itemize}
   \item[(1)] To have a method of preparing the observed system in any one of a set of well-defined, reproducible initial states.
   \item[(2)] To have in the formalism the class of reversible isentropic processes of classical thermodynamics and the notion of temperature.
\end{itemize}

The first requirement is often overlooked.  The processes of current interest begin with a system in thermal equilibrium and a work cycle brings it out of equilibrium.  The initial equilibrium state is usually assumed without question. Lindblad warns a possible tautology and writes "The introduction of \textit{a priori} irreversible preparation procedure in a formalism which has some ambition to explain irreversibility on the basis of reversible microscopic dynamics may seem to constitute a vicious circle." and "For this reason the state preparing procedures will be defined by a class of highly idealized heat baths which can be specified in a microscopic way." (page 33)

Lindblad remarks that it is difficult to develop a general theory of relaxation to a thermal state.  However, he was optimistic that finding a single working model is good enough, in his own words, "it can be argued that it is sufficient to have one microscopic model with this property even if it is rather artificial." (page 37)

The microscopic models of heat baths he considered are infinitely large quasifree\footnote{In the current context, it is sufficient to assume that the correlation functions of all orders are given by the products of the second order correlation functions or otherwise vanish.} boson or fermion systems. Their temporal correlation functions decay rapidly so that the heat baths effectively remain in thermal equilibrium (a KMS state) within the time scale of coupling between the system and the heat baths.  This allows the weak coupling limit (WCL), which in turn allows us to use the Lindblad equation for the time evolution of the system.   As long as various time scales, including the one associated with the work cycle, satisfy the given conditions 
as developed by Lindblad in Appendix A of the book, see also \cite{Davies1976,Spohn1980}, the microscopic models work as ideal heat baths.  The same models are widely  used up to the present day\cite{Breuer2007}.

In most cases studied in the book, a target process begins with a Gibbs state with a Hamiltonian $H \in F$ and the system is driven to a non-equilibrium state by a work cycle.  It is convenient to define the set $E(F)$ of states which are F-accessible from the Gibbs states.  "The set contains all the experimentally available non-equilibrium states.  It is for these states that we have to define the entropy function."

In addition to heat baths, the system interacts with the rest of the world (X) but very weakly, much weaker than the interaction with the heat baths, even under WCL. We want to make sure that S can be thermalized with R, but not with X. Otherwise, the entire universe falls into  the "heat death".  To prevent this, energy exchange between S and X must be negligible. 
Lindblad briefly discusses the necessary conditions in this chapter.  The effects of such weak perturbation on non-equilibrium states are discussed again in Chapter 8.

\subsection{Chapter 5 "Reversible processes"}
\label{sec:synopsis-5}

In this chapter, Lindblad analyzes macroscopically reversible processes from the microscopic point of view using the definitions introduced in the previous chapters (sections in above), namely the F-accessible work cycle, information entropy and the Gibbs state from Chapter 2 (Sec.~\ref{sec:synopsis-2}), 
and ideal heat baths and the weak coupling limit from Chapter 4 (Sec.~\ref{sec:synopsis-4}) 
As discussed above in Section~\ref{sec:synopsis-2}, if F is large enough, any work cycle applied to a closed system is microscopically reversible. The focus of this chapter is conversely on thermodynamic reversibility for infinitely large systems (thermodynamic limit) where F can be a finite set. In standard thermodynamics, quasi-static processes where the system and the heat baths remain in thermal equilibrium are considered reversible in general. 
The goal Lindblad sets for himself is to confirm it from a microscopic picture.

Lindblad considers the following thermodynamic process. The whole system consists of a finite subsystem S and a reservoir R. Throughout the chapter, S+R is treated as a single closed system, and density operators $\rho$ express joint states (without index S+R).  S and R are initially in thermal equilibrium $\rho_0$ at an inverse temperature $\beta$. When a work cycle $\gamma$ is applied to S, the initial state $\rho_0$ is transformed to a new joint state $\rho_1$.  After the work cycle is completed, the whole system is relaxes to a new thermal equilibrium $\rho_1^\prime$.  This process is not necessarily reversible and $\rho_1$ can be non-equilibrium.  For it to be reversible, $\rho_1$ and $\rho_1^\prime$ must satisfy certain conditions,
which Lindblad proceeds to work out.

Standard thermodynamics says that if the process is reversible, $\Delta S^R = Q/T$ where $Q$ is reversible heat. 
Recall 
that information entropy corresponds to thermodynamic entropy if the system is in thermal equilibrium. Since S returns
back to the original state,  $\Delta S^S = 0$ and $Q+W=0$ (the first law of thermodynamics).  The reversibility condition is then expressed as
\begin{equation}\label{eq:std-formula}
    \Delta S^R = \frac{Q}{T} \quad \Rightarrow \quad \Delta S_I^R = -\beta W(\gamma,\rho(\beta,H)) 
\end{equation}
which Lindblad calls the "standard formula".

The aim of this chapter can now be said to be to
verify Eq. \eqref{eq:std-formula} for the current process. If we naively think that the system
returns 
back to the original state
and that the state of R does not
change we find that
$\rho_1^\prime=\rho_0$, 
and consequently that
$\Delta S_I^{S+R} = S_I(\rho_1^\prime) - S_I(\rho_0)$ vanishes, in contradiction to the standard formula. In the thermodynamic limit, 
we however have
$\infty^\prime-\infty$ which can be finite. 
Lindblad avoids 
this trap by carefully evaluating $\Delta S_I^{S+R}$ in the thermodynamic limit.

We follow the Lindblad's argument, which is more involved than it may 
appear at first sight.  By substituting $\rho(\beta,H)=\rho(s)=\rho_0$ and $\rho(t)=\rho_1$ in Eq. \eqref{eq:work-S} and using $S_I(\rho_0|\rho_0)=0$, it follows that
\begin{equation}\label{eq:-W=S(r1|r0)}
   \beta W =  - S_I^{S+R}(\rho_1|\rho_0),
\end{equation}
which is exact without the thermodynamic limit. In the following, Lindblad  shows that $\Delta S_I^{S+R} = S_I(\rho_1|\rho_0)$ in the thermodynamic limit.  The proof is rather complicated since the right hand side does not depend on the final state.  Hence, we must find the relation between $\rho_1$ and $\rho_1^\prime$.

First of all, temperature must remain the same, or otherwise Eq. \eqref{eq:std-formula} would not make sense.
  On the other hand, the work cycle injects some energy (negative work) and thus 
in a large but finite 
total system S+R (inverse) temperature would
change to $\beta'$ corresponding to final energy $\Tr{H\rho_0} - W$,
see Section~\ref{sec:synopsis-3} 
Injecting a physical 
principle into the argument
Lindblad assumes that when R is infinitely large, its heat capacity diverges, and thus $\beta^\prime = \beta$.  However, this does not mean that $S_I(\rho_0)=S_I(\rho_1^\prime)$ as discussed above. 

To find the relation between $\rho_0$ and $\rho_1^\prime$, Lindblad calculates the asymmetry between $S_I(\rho_1^\prime|\rho_0)$ and $S_I(\rho_0|\rho_1^\prime)$. Substituting $\rho(\beta,H)=\rho_0$ and $\rho=\rho_1^\prime$ in Eq. \eqref{eq:thermo_stability} and $\rho(\beta^\prime,H)=\rho_1^\prime$ and $\rho=\rho_0$ also in the same equation, we obtain two  relative entropies
\begin{subequations}\label{eq:two-beta}
    \begin{equation}\label{eq:two-beta1}
		S_I(\rho_1^\prime|\rho_0) = S_I(\rho_0) - S_I(\rho_1^\prime) - \beta W.
	\end{equation}
    \begin{equation}\label{eq:two-beta2}
        S_I(\rho_0|\rho_1^\prime) = S_I(\rho_1^\prime) - S_I(\rho_0) + \beta^\prime W
    \end{equation}
\end{subequations}
By adding the two equations, it follows that
\begin{equation}\label{eq:S(r1'|r0)}
	S_I(\rho_0|\rho_1^\prime) + S_I(\rho_0|\rho_1^\prime) = (\beta^\prime - \beta) W.
\end{equation}
The right hand side vanishes in the thermodynamic limit and we obtain\footnote{\label{fn:relativeS}Recall the basic property of relative entropy $S_I(\rho|\sigma) \ge 0$ and  $S_I(\rho|\sigma)=0$ if and only if $\rho=\sigma$.}
\begin{equation}\label{eq:S(r1'|r0)=0}
	\lim_{R \rightarrow \infty} S_I(\rho_1^\prime|\rho_0) = \lim_{R \rightarrow \infty} S_I(\rho_0|\rho_1^\prime) = 0.
\end{equation}
which provides an important relation between $\rho_0$ and $\rho_1^\prime$.  Again, this result should not be interpreted as $\rho_1^\prime=\rho_0$.

Next, Lindblad 
searches for
a relation between $\rho_1$ and $\rho_1^\prime$ by taking the limit of Eq. \eqref{eq:-W=S(r1|r0)}.
\begin{eqnarray}\label{eq:-W=S(r1|r1')}
	-\beta W &=& S_I(\rho_1|\rho_0) = \Tr(\rho_1 \ln \rho_1) - \Tr(\rho_1 \ln \rho_0) \nonumber \\
	&=& \Tr(\rho_1 \ln \rho_1) + \beta \Tr(\rho_1 H) + \beta \ln Z \nonumber \\
	&\xrightarrow{\beta \rightarrow \beta^\prime}&   \Tr(\rho_1 \ln \rho_1) + \beta^\prime \Tr(\rho_1^\prime H) + \beta^\prime \ln Z^\prime \nonumber \\
	&=& S_I(\rho_1|\rho_1^\prime)
\end{eqnarray}
where we used $\Tr(\rho_1 H) = \Tr(\rho_1^\prime H)$.  This result can be obtained also by taking the limit of the chain rule\footnote{For any $\rho$, the chain rule between the Gibbs state $G_1$  and $G_2$ is given by $S_I(G_1|G_2) = S_I(G_1|\rho) + S_I(\rho|G_2)$.} $S_I(\rho_1|\rho_0) = S_I(\rho_1|\rho_1^\prime) + S_I(\rho_1^\prime|\rho_0)$. 
Hence, we found a missing relation between $\rho_1$ and $\rho_1^\prime$ in the thermodynamic limit, that is $S_I(\rho_1|\rho_0) = S_I (\rho_1|\rho_1^\prime)$.

Based on Eqs. \eqref{eq:-W=S(r1|r0)} and \eqref{eq:-W=S(r1|r1')}, Lindblad 
posits, without justification, that
\begin{equation}\label{eq:redefineS}
  -\beta W = S_I(\rho_1|\rho_1^\prime) \equiv \Delta S_I^{S+R} .
\end{equation}
At first sight, this looks strange since the entropy change  does not depend on the initial state $\rho_0$.  It can be verified as follows.
Using Eq. \eqref{eq:thermo_stability} again with $\rho(\beta,H)=\rho_1^\prime$ and $\rho=\rho_0$ and $W=0$ (no work during the relaxation period), it follows that
\begin{equation}\label{eq:S(r1')-S(r1)}
	S_I(\rho_1|\rho_1^\prime) = S_I(\rho_1^\prime)-S_I(\rho_1) .
\end{equation}
Now adding Eq. \eqref{eq:S(r1')-S(r1)} to Eq. \eqref{eq:two-beta1} and using Eqs. \eqref{eq:S(r1'|r0)=0} and \eqref{eq:-W=S(r1|r1')}, we obtain
\begin{eqnarray}
S_I(\rho_1^\prime|\rho_0) + S_I(\rho_1|\rho_1^\prime) &=& S_I(\rho_0) - S_I(\rho_1) - \beta W  \nonumber \\
\xrightarrow{R\rightarrow \infty}  \qquad -\beta W &=&  S_I(\rho_0) - S_I(\rho_1) - \beta W  \nonumber \\
\rightarrow \qquad S_I(\rho_0) &=& S_I(\rho_1). \label{eq:S(r1)=S(r0)}
\end{eqnarray}
which means the entropy does not change during the work cycle. Replacing $S_I(\rho_1)$ with $S_I(\rho_0)$ in  Eq. \eqref{eq:S(r1')-S(r1)}, at last we obtain the correct thermodynamic limit
\begin{equation}\label{eq:S(r1|r1')=DS}
	\Delta S_I^{S+R}  = S_I(\rho_1^\prime)-S_I(\rho_0) = S_I(\rho_1|\rho_1^\prime) .
\end{equation}
which justifies Eq. \eqref{eq:redefineS}.

Combining Eqs. \eqref{eq:-W=S(r1|r1')} and \eqref{eq:S(r1|r1')=DS}, we conclude that the the change of entropy in the thermodynamic limit is given by
\begin{equation}\label{eq:DS-final}
	\Delta S_I^{S+R} = S_I(\rho_1|\rho_1^\prime) = S_I(\rho_1|\rho_0) = - \beta W
\end{equation}
Since under the assumption of a weak coupling limit (WCL) the correlation 
between S and R can be neglected, we have $\Delta S_I^{S+R}=\Delta S_I^S + \Delta S_I^R$.  Furthermore,  $\Delta S_I^S=0$ since S returns to the initial state, see Section \ref{sec:synopsis-4}.  Then, Eq. \eqref{eq:DS-final} is equivalent to the standard formula \eqref{eq:std-formula}.

It is admittedly easy to lose one's way in the above derivation due to the
apparently counter-intuititive relations obtained in the thermodynamic limit.  Lindblad provides a physical reasoning why the entropy increases despite 
the system returning back to thermal equilibrium at the same temperature by introducing the "quasilocal" description of the system in the thermodynamic limit.  The whole system is divided into many finite subspaces $\{\Lambda\}$.  If there is local relaxation back to equilibrium, then for any subspace $\Lambda$, the reduced densities satisfy  $\rho_{0,\Lambda} = \rho_{1,\Lambda} = \rho_{1,\Lambda}^\prime$.  However, the global states $\rho_0$, $\rho_1$ and $\rho_1^\prime$ do not have to be the same if there is correlation among the subspaces.  Equation \eqref{eq:DS-final} clearly shows that $\rho_1$ differs from $\rho_0$ and $\rho_1^\prime$, indicating that the global correlation present in $\rho_1$ is responsible for the change of the entropy.

When the process is not reversible, $\rho_1$ is a non-equilibrium state and irreversible entropy is produced. 
One example of a recent contribution showing that the origin of the entropy production is mutual information formed inside the heat bath, which seems consistent with Lindblad's idea in this chapter,
is~\cite{Ptaszynski2019}.

\subsection{Chapter 6 "Closed finite systems"}
\label{sec:synopsis-6}
In Chapter 6, Lindblad discusses to what extent irreversibility can be defined for a finite quantum system with Hamiltonian dynamics in terms of F-accessible  work cycles defined in Chapter 2. In particular, he defines an entropy function for closed system, which forms the basis of the final definition of non-equilibrium entropy developed in Chapter 10.

From Eq. \eqref{eq:subset-Omega-mu}, irreversibility is defined by
\begin{equation}\label{eq:irreversibility-Omega}
   \Omega(\rho(t),F) \subset \Omega(\rho(s),F); \quad t>s.
\end{equation}
When F contains all self-adjoint operators, then the two sets coincide and the processes are reversible. Choosing an origin of work $H \in F$ and an initial state $\rho$, Lindblad introduces a greatest lower bound of accessible energy:
\begin{equation}\label{eq:irreversibility-Q}
Q(\rho;F,H) =  \inf\{\mu(H), \mu \in \Omega(\rho,F)\}
\end{equation}
This is the lowest (expected) energy that the system can reach by the available work cycles.  From the irreversibility condition \eqref{eq:irreversibility-Omega}, it follows that $Q$ only grows during irreversible processes. The passive state $\rho$ can be defined as $Q(\rho;F,H) = \Tr(\rho H)$, \textit{i.e.} the (expected) energy of a passive state with respect to $F$ cannot be decreased by any work cycle based on $F$. Lindblad further introduces the \emph{availability function}
\begin{equation}\label{eq:irreversibility-A}
A(\rho;F,H)= \Tr[\rho H] - Q(\rho;F,H)
\end{equation}
which is the supremum of the work that can be be obtained from the work cycles in $F$, starting in state $\rho$ and Hamiltonian $H$. A work cycle $\hat\gamma$ defined by $A(\rho;F,H) = W(\hat\gamma)$ extracts all available work. The infimum in Eq.~\ref{eq:irreversibility-Q} means that $\hat\gamma$ may not exist as a proper work cycle, but that $A(\rho;F,H)$ can be arbitrarily well approximated by actual work cycles. Both greatest lowest accessible energy and the availability function are used to construct non-equilibrium entropy in Chapter 10.

The Boltzmann's H theorem was criticized based on the two aspects of microscopic dynamics, the reversibility by Loschmidt and the recurrence by Zermelo\cite{Brown2009,Mackey1993}.
Lindblad expected similar criticisms and defended his formalism against them.  Since the F-accessibility introduces the irreversibility, the Loschmidt's paradox does not apply to his formalism. 
Lindblad further comments that "for a classical system with the mixing property the Poincar\,{e} recurrence theorem does not really present a difficulty, ...," and  the recurrence paradox of Zermelo can be mitigated for such systems~\cite{Mackey1993}. However, Lindblad admits that "The recurence paradox of Zermelo is not so easy to dispose of for a finite quantum system."

In the second section of Chapter 6, Lindblad address the issue of recurrence. He first considers the possibility of arriving at (expected) energy $E$ (less than $Q(\rho;F,H)$) in a finite time by some work cycle in the set
$\Gamma(F)$, and defines $\tau(\rho(t),E)-t$ to be the infimum of such times. Lindblad shows that
\begin{equation}
\tau(\rho(s),E)\leq \tau(\rho(t),E)\quad\hbox{for all $s\leq t$}.
\end{equation}
from which he advances the hypothesis that for many complex systems
\begin{equation}
\tau(\rho(t),E)- \tau(\rho(s),E) >> t-s
\end{equation}
and observes that if this is so, evolution has made it harder to access the same energy. Lindblad interprets this effect as the "origin of irreversibility in finite system". Note that this irreversibility of finite closed systems in the sense of Lindblad is not absolute, but depends on $F$. It is certainly a very original idea which has not have much continuation in the later literature.

After defining irreversibility in terms of energy and justifying it against Loschmidt and Zermelo paradoxes, Lindblad defines an entropy function for all F-accessible non-equilibrium states $\rho \in E(F)$ with a given initial Hamiltonian $H \in F$. This definition relies on existence of a function $\beta(u,H)$ (introduced by Lindblad in Section~3.b) which is the inverse temperature at which the expected energy under the Gibbs ensemble based on $H$ equals $u$. We recall the equilibrium statistical mechanics expression for entropy as function of energy by thermodynamic integration, $S(E)=\int_0^E \beta(u,H) du$. Lindblad defines \textit{thermodynamic entropy for a work cycle $\gamma$} as
\begin{equation}
\label{eq:S-gamma}
S(\rho;\gamma) = \int_0^E \beta(u,H) du \quad E=\Tr[\rho H]-W(\gamma,\rho)
\end{equation}
and thermodynamic entropy relative to $\rho$ and for the set $\Gamma(F)$ (or the mobility semigroup $T(F)$) as
\begin{equation}
\label{eq:S-F}
S(\rho;F) = \inf \{ S(\rho;\gamma); \gamma \in \Gamma(F) \}
\end{equation}

This entropy has the following properties:
\begin{itemize}
   \item[(1)]  $S_I(\rho) \le S(\rho;F)$  (Equality holds if $\rho$ is a Gibbs state.)
   \item[(2)]  $S(\rho(s);F) \le S(\rho(t);F)$  for all $s<t$
   \item[(3)]  $S(\rho,F) \le S(\rho,F')$ for $F \subseteq F'$.
   \item[(4)]  $\sum_k \lambda_k S(\rho_k;F) \le S\left(\sum_k \lambda_k \rho_k;F\right)$,  for $\rho_k \in E(F)$,  $\lambda_k>0$,  $\sum_k \lambda_k$.
\end{itemize}
The non-decreasing function (2) and convexity (4) are required for any entropy functions.  The properties (1) and (3) need some remarks. (1) follows from that von Neumann-entropy is conserved during unitary evolution (in the chapter under review all evolutions are unitary). Whatever $\rho'$ is reached by the work cycle, its von Neumann entropy cannot be
larger than entropy of the Gibbs state with the same expected energy, as maximizing entropy under constraints is a defining property of Gibbs states. Hence $S_I(\rho)$ is less than or equal to $S(\rho;\gamma)$ in Eq.~\eqref{eq:S-gamma} for any $\gamma$, and therefore in particular less than or equal to their infimum, which is $S(\rho;F)$ in Eq.~\eqref{eq:S-F}. For the largest $F$ that contains all self-adjoint operators,  we have the minimal entropy function from property (3). We can show that it coincides with $S_I(\rho)$.  Then, $\rho$ must be a Gibbs state from (1). As discussed earlier, the largest F allows reversible processes.  Hence, the entropy of reversible processes is $S_I(\rho)$ as expected.

The second law of thermodynamics under unitary evolution is still a major issue.  The eigenthermalization hypothesis (ETH) has been quite successful to explain thermal states of closed systems~\cite{Deutsch1991}.  However, the information entropy is conserved by unitary transformation and thus cannot be used to show the second law as Lindblad stated in Chapter 3 (See also~\cite{Deutsch2018}).  It has been proposed to use the classical Shannon entropy in the energy eiegenbasis, so-called \emph{diagonal entropy} (d-entropy)~\cite{Ikeda2015}.  
The relation between the Lindblad's approach and d-entropy, if any, however remains to be established.

\subsection{Chapter 7 "Open systems"}
\label{sec:synopsis-7}

Lindblad starts this chapter by recalling a system $S$ weakly coupled to a heat bath $R(\beta)$ at inverse temperature $\beta$, and states that the objective of the open system approach he is to discuss is to "treat the dynamics and thermodynamics of $S+R$ using only the reduced description given by the partial state $\rho_S$ of $S$ and the parameter $\beta$ of $R$" (page 60, lines 12-15).

From the modern point of view it is clear that the above is not an entirely achievable goal. Thermodynamic functions such as work are in general functionals of the open quantum system evolution, similar but not identical to the open quantum state \cite{Aurell2020}. Lindblad in facts remarks (page 60, lines 17-19) that he is here introducing an extra requirement.

The chapter should perhaps best be seen as the beginning of a discussion of the limitations of modelling open quantum systems as quantum Markov processes, which at the time was but very incompletely understood. The chapter is also an entry point to a more technical appendix A (16 pages) where Lindblad discusses these issues in more detail. The appendix begins with a very clear and succinct summary of Lindblad's then fairly recent theory of quantum Markov processes.

\subsection{Chapter 8 "External Perturbations"}\label{sec:synopsis-8}

In Chapter 8 Lindblad discusses a possible mathematical model of \emph{essentially isolated systems} introduced in Chapter 3. The question was how the information entropy $S_\textsc{i}$ increases in an isolated system.  We know that it remains constant during any unitary evolution. In a thermodynamics context, however, an isolated system does not exchange energy with its environment but the interaction between them does not have to be absent.  The interaction can change the information entropy of the system without energy exchange if the correlation between them changes.\footnote{As an example, consider decoherence induced by the environment.  The decoherence changes entropy but energy exchange is not required.  See Ref.~\cite{Schlosshauser2007}.} Such environments (X) are introduced in Chapter 4. They interact with the system much weaker than heat baths (R).  Because the stability of equilibrium states,  the weak perturbation exerted by X will not affect the initial equilibrium state (see Chapter 4).  However, non-equilbrium states will not exhibit such nice properties and there could be a significant influence on the evolution of non-equilibrium states increasing its entropy.  In Chapter 4, the desired properties of R is clearly defined and thus we can make a mathematical model with the desired properties.  On the other hand,  X is completely out of control except that the coupling between S and X is so weak that S+X will not reach equilibrium.  X  may not have even a definite temperature.  The lack of specificity makes it difficult to develop a mathematical model.

The key question is how unstable the evolution of non-equilibrium states is under the perturbation by X.  Lindblad sees a certain similarity between the instability caused by the perturbation and by the classical chaos.  However, no quantum version of the mixing property and the Kolmogorov-Sinai entropy is known. To extend the idea of classical Hamiltonian chaos to quantum dynamics, Lindblad first tried to quantify the instability using information entropy available for both classical and quantum systems. 
This is developed in Appendix B "Sensitivity of Hyperbolic Motion".  Without a rigorous mathematical theory, the idea is phenomenologically extended to quantum systems based on the energy-time uncertainty relation. The concept of quantum chaos was still in its infancy at the time the book was published.  However, 
Lindblad's discussion seems to be in line with modern quantum chaos theory.
He concluded that the entropy function introduced in Chapter 6 only very weakly depends on the coupling strength and most likely the instability is physically not significant.

\subsection{Chapter 9 "Thermodynamic limit"}
\label{sec:synopsis-9}
This chapter lies a bit outside the main focus of this review. Stochastic thermodynamics and its quantum analogue quantum thermodynamics is (mainly) concerned with systems that may interact with large reservoirs, but which are not themselves in the thermodynamic limit (limit of unlimited number of degrees of freedom).

\subsection{Chapter 10 "Thermodynamic entropy"}
\label{sec:synopsis-10}

In Chapters 2-9, Lindblad introduced components necessary to construct a general definition of \emph{thermodynamic entropy} for non-equilibrium states.  In Chapter 10, he 
ties these strands together and posits a thermodynamic entropy function for non-equilibrium states and presents its properties.   Consider a set of thermodynamic processes P=P(F) which includes all processes investigated in the previous chapters, both Hamiltonian dynamics for closed systems, and Markovian dynamics under the weak coupling limit. Interaction with multiple heat baths with different temperature are also included. The main goal is to find an entropy function $S(\rho;P)$ for any state $\rho$ accessible by $P$.   To do so, Lindblad considers a hypothetical process driven by a work cycle $\gamma$ which brings an arbitrary state $\rho$ to a Gibbs state $\rho(\gamma) \in G(F)$.  Then, he defines thermodynamic entropy, or "P-entropy", as
\begin{equation}\label{eq:central-eq}
   S(\rho;P) = \inf_\gamma \{S_I(\rho(\gamma)) + \sum_R \Delta S^R(\rho;\gamma) \}
\end{equation}
where the change of entropy in heat bath R is $\Delta S^R (\rho;\gamma)=\beta(R) \Delta E^R$ as derived in Chapter 5, and
the infimum is taken over all $\gamma \in \Gamma(F)$ that takes $\rho$ to a Gibbs state.

The definition has a simple interpretation based on the second law of thermodynamics. Recall that entropy 
changes can be classified as reversible and irreversible. Heat $Q$ 
can also be classified as reversible and irreversible.  Then, the total change in entropy during a work cycle $\gamma$ can be written as 
$\Delta S(\gamma) = S_I(\rho(\gamma)) - S(\rho;P) = -\sum_R Q^R(\gamma)/T + \Sigma(\gamma)$ where $\Sigma(\gamma)$ is the irreversible entropy production. The second law demands that $ 0 \le \Sigma(\gamma) = S_I(\rho(\gamma)) - S(\rho;P) + \sum_R Q^R(\gamma)/T$.  Thus, $S(\rho;P) \le S_I(\rho(\gamma)) +  \sum_R \Delta S_I^R(\rho;\gamma);  \forall \gamma \in \Gamma(F)$.  Here we used $\sum_R Q^R(\gamma)/T =  \sum_R \Delta S_I^R(\rho;\gamma)$ (see Sec. \ref{sec:synopsis-5}).  The infimum in Eq. \eqref{eq:central-eq} is sufficient to satisfy the second law. In other words, the infimum is equivalent to optimizing the entropy production with respect to  $\gamma$.  
Lindblad appears to have
been the first to consider a 
definition of non-equilibrium entropy
of this type. The concept is quite closely
related to much more recent research, and we
will return to this connection in
Section~\ref{sec:fore-runner} 

Lindblad shows that the P-entropy \eqref{eq:central-eq} satisfies a set of following inequalities:
\begin{itemize}
   \item[(1)]  For all $\rho \in E(F)$, $S_I(\rho) \le S(\rho;P)$ and equaility holds for $\rho \in G(F)$.
   \item[(2)]  For all time interval $s \le t$, $S(\rho(s);P) \le S(\rho;P) + \sum_R \Delta S^R(s,t)$.
   \item[(3)]  For $P' \subseteq  P$, $S(\rho;P) \le S(\rho,P')$.    If $P$ is large enough, the inequality implies $S(\rho;P) = S_I(\rho)$.  (See Chapter 6 for microscopic reversibility for F containing all self-adjoint operators.)
   \item[(4)]  For any set $\{\rho_k \in E(F), \lambda_k \ge 0; \sum_k \lambda_k = 1\}$, $\sum_k \lambda S(\rho_k;P) \le S(\sum_k \lambda k \rho_k;P)$.
   \item[(5)]  For independent subsystems $S_1$ and $S_2$  (not coupled and driven by independent Hamiltonians), $S(\rho;P) = S(\rho_1;P_1) + S(\rho_2;P_2)$.
\end{itemize}

One of important consequences from the above set of inequalities is a precise description of irreversibility.  Suppose that an initial equilibrium state $\rho_0$ is driven to a  state $\rho$. We find from (1) and (2) that if the process is an optimal adiabatic process then  $S(\rho;P) = S_I(\rho)$ and for any irreversible processes, we have
\begin{equation}\label{eq:essential_irreversility}
   S(\rho;P) \ge S_I(\rho).
\end{equation}
When the mobility semigroup T(F) is large enough, i.e., F contains all self-adjoint operators,  $S(\rho;P) \simeq S_I(\rho)$ as property (3) indicates.  Then, the irreversibility is essentially due to the increase in the I-entropy induced by the external perturbation. (See Chapters 7 and 8.)  On the other hand, when $T(F)$ is limited, the P-entropy significantly deviates from the I-entropy due to the limited intrinsic Hamiltonian dynamics. The irreversility is caused by the limitation of the F-accessible states. (See Chapter 6.)

Next, Lindblad discusses \emph{approach to equilibrium} using the P-entropy.  To do so, he introduces a P-relative entropy\footnote{Not confused with the standard relative entropy.}
\begin{equation}
   S\left(\rho|\rho(\beta,H)\right)\equiv S(\beta,H)-S(\rho;P)+\beta\left[\ev{H}_\rho - \ev{H}_{\rho(\beta;H)}\right].
\end{equation}
as a distance between $\rho$ and the Gipps state similar to Eq. \eqref{eq:thermo_stability}.
Like a regular relative entropy, it vanishes if and only if $S(\rho;P)=S(\beta;H)$ and $\ev{H}_\rho = \ev{H}_{\rho(\beta;H)}$.  Furthermore,  $S(\rho(t)|\rho(\beta,H);P)$ is a non-increasing function of time $t$. Then, Lindblad introduces the concept of P-equilibrium, the closest of P-accessible Gibbs states. When the displacement
\begin{equation}
   d(\rho|G(F)) = \inf_\mu \{S(\rho|\mu;P); \mu \in G(F)\}
\end{equation}
vanishes, $\rho$ is the P-equilibrium state.  Since $d$ is a non-increasing function, the system tends to approach the P-equilibrium.

\subsection{Chapter 11 "Measurements, entropy and work"}
\label{sec:synopsis-11}

In this last chapter  Lindblad tackles the issue of quantum measurement.  Based on the standard interpretation of quantum mechanics, quantum measurement transforms the state of measured system to a different state and the transition is irreversible. There is no 
universally agreed upon
dynamical process that describes the quantum measurement.  Lindblad considers quantum measurement as non-equilibrium thermodynamics processes taking place between the system (S) and the measurement device (M).  The mathematical part is based on a series of papers published earlier\cite{Lindblad1972,Lindblad1973,Lindblad1974}.  The physical interpretation of the quantum measurement is not conclusive within the book. Part of his interpretation
was clarified in a paper published 28 years later~\cite{Lindblad2011}.

Lindblad first shows that the classical information obtained by the measurement has a lower bound.  Consider a measurement of system in a state $\rho$.  The outcome $\rho_k$ is obtained with probability $p_k$.  If $\rho_k$ is a pure state $\dyad{k}$, this is a standard projective measurement.  Otherwise, it is \emph{non-efficient} measurement.
Lindblad further assumes that the measurement is \emph{non-destructive} and thus $\rho=\sum_k p_k \rho_k$\footnote{
We do not impute to Lindblad to the position that the destructive nature of quantum measurement is not essential for the discussion.  To the contrary, it is essential, especially when there is back action.}. Before the measurement, the system carries an amount $S_I(\rho)$ of information. When the measurement outcome 
has been recorded, the amount of classical information obtained by the observer is given by Shannon entropy $I\{p_k\} = - \sum_k p_k \ln p_k$: for non-efficient measurement, some information remains in the 
system, on average $\sum_k p_k S_I(\rho_k)$.  If the outcome is not recorded, the amount of information in the system remains $S_I(\rho$) since the measurement is assumed to be non-destructive.  Using convexity and sub-additivity of information entropy, Lindblad shows that
\begin{equation}\label{eq:information_gain}
	0  \le S_I(\rho) - \sum_k p_k S_I(\rho_k) \le I\{p_k\}
\end{equation}
which is a well-known inequality in quantum information theory~\cite{Nielsen2000}.

The above inequality does not take into account the change in the state of M nor the correlation between S amd M. To emphasize their importance, Lindblad invokes the Maxwell demon.  Suppose that the system is in a Gibbs state which is passive with respect to P-work cycle.  After measurement, the system jumps to $\rho_k$ which is no longer passive.  Hence, it is possible to extract work out of it with another P-work cycle, which seemingly violates the second law of thermodynamics.  He recognized that this is a quantum version of the Maxwell demon, the measurement device being the demon.  

Lindblad mentions the 
Szilard engine as example without detailed analysis
of the quantum version
which was 
published much later~\cite{Kim2011,Sagawa2009}.
Lindblad then analyzes the information in the joint system S+M and derives the upper bound of the classical information gain
\begin{equation}
    I\{p_k\} \le C_{12}(\rho^\prime_{12})
\end{equation}
where $C_{12}$ is the mutual information in the final state of S+M, $\rho^\prime_{12}$.
Then, Lindblad goes on to evaluate the energy and heat (information) exchange between S and M.

Throughout the chapter, Lindblad uses
the P-entropy to explain the thermodynamics of quantum measurement. This suggests
that in his mind the underlying dynamics of quantum measurement is a quantum Markov processes.  Lindblad points out that non-equilibrium processes and quantum measurements share a similar irreversibility.  In his much later paper~\cite{Lindblad2011}, he suggested that the irreversibility in quantum measurement is due to ever expanding quantum entanglement in the measurement device similar to the global correlation discussed above in Chapter 5. 
In both~\cite{Lindblad2011} and in his last
unpublished paper~\cite{Lindblad2022},
Lindblad cites in this connection the quantum Darwinism of Zurek~\cite{Zurek2009}.

\subsection{Chapter 12 "Other approaches"}
\label{sec:synopsis-12}
In this last chapter of the book Lindblad summarizes and discusses critically other approaches to non-equilibrium entropy actively investigated at the time when the monograph was written. From the perspective of this review of Lindblad's work, the main interest is what other approaches he had in mind, and the role of these other approaches today, \textit{i.e.} 40 years later.

Lindblad begins by discussing the work of the Brussels group led by Prigogine, which is still actively considered to this day, and which has been presented in several books and monographs, for example, see Ref.~\cite{Prigogine1980}. The essence of the Brussels approach is that classical chaos leads to growth of uncertainty, which is then extended to a postulate of intrinsic irreversibility. The exact nature of this may perhaps be said not to have crystalized in one precise form in the presentation of the Brussels group and their supporters, critics and detractors, but in the current context this is immaterial. Lindblad's main argument against the Brussels approach is that conservative (reversible) evolution exists in nature, at least to very good approximation, and that this is difficult to reconcile with any kind of intrinsic irreversibility. A later critique against Prigogine's ideas can be found in the writings of Bricmont~\cite{Bricmont1995}.

The second approach discussed by Lindblad, which he attributes to Penrose and Goldstein \cite{Goldstein1981}, aims to connect entropy of a state $S$ (thermodynamic or other entropy) to the Kolmogorov-Sinai entropy $h$ describing the rate of gain of information of a classical dynamical system. Lindblad remarks that this introduces a characteristic time as $t=\frac{S}{h}$ such that $S$ is produced in a time $t$, and that beyond that time information as measured by Kolmogorov-Sinai entropy can increase indefinitely.

The third approach is the \textit{subjective} interpretation of entropy as measuring ignorance: Lindblad also calls this the \textit{anthromorphic} 
\footnote{Likely a reference to \cite{Jaynes1965} (Section VI), cited by Lindblad further on in this Chapter. The phrase 
"Entropy is an anthropomorphic concept" is in~\cite{Jaynes1965} attributed to E.P. Wigner.} approach. In this approach it is postulated that information-theoretic entropy $-\sum_k p_k \log p_k$ -- of practically any probability distribution -- is to maximized given constraints on the probabilities
$p_k$ representing one's prior knowledge. Jaynes~\cite{Jaynes1957} introduced this as the \textit{maximum-entropy method} in 1957. As Lindblad remarks maximum-entropy naturally appears as a form of statistical inference, and as such it has remained popular to this day. Another early proponent of thermodynamics (and entropy) as a consequence and/or a form of statistical inference was Mandelbrot~\cite{Mandelbrot1962}.

One prominent recent application of these ideas is the use of Gibbs-Boltzmann distributions to recover interactions from large-scale biological data. This undisputed success of maximum-entropy inference is
in a view which we subscribe to
due not to "knowledge out of ignorance", but because many natural distributions are of the Gibbs-Boltzmann distribution type, where maximum-entropy is equivalent to maximum-likelihood inference \cite{Aurell2016,Dichio2023}, see also Bricmont~\cite{Bricmont2019}. Many in the community working on this class of problems are however of the opposite opinion, see \textit{e.g.} \cite{vanNimwegen2016}. In thermodynamics in a more strict sense the approach of Jaynes and Mandelbrot has in any case largely fallen out of favor in recent years, for a critical discussion, see \cite{Auletta2017}.

The final approach discussed by Lindblad is that of "rational thermodynamics", referring to two papers in the mathematical journal \textit{Archive for Rational Mechanics and Analysis}, by respectively Day \cite{Day1968} and Coleman and Owen \cite{Coleman1974}. Although the second of these papers is rather well cited, this approach has not had a large impact in the later physical literature.

\section{Was Lindblad a forerunner of modern non-equilibrium thermodynamics?}
\label{sec:fore-runner}
Lindblad introduced many unique ideas in his book. We will now 
survey selected recent developments related to his pioneering contribution.

\vspace{2ex}
\noindent\textbf{Open quantum systems and quantum master equations}: 
Standard quantum mechanics textbooks assume that quantum objects are completely isolated from the rest of the world. Such 
an
idealization also played an important role when classical mechanics was developed.  However, the influence of the environment on quantum systems is much more significant than on classical systems.  For example, the quantum coherence in the system can be destroyed by the environment even without energy exchange~\cite{Schlosshauser2007}.
The theory of open quantum systems 
is essential for the development of quantum technologies. 

The investigation of open quantum systems were well underway even before  Lindblad's book was published. See for example the Davies's book 
"Quantum Theory of Open Systems"~\cite{Davies1976}, published in 1976. The goal is to derive equations of motion for \emph{open quantum systems} in contact with quantum heat baths.  Due to dissipation, the resulting time evolution is not unitary.  It is known as \emph{completely positive dynamical semigroup}~\cite{Alicki2007}.  In his 
famous 1976 paper~\cite{Lindblad1976},  Lindblad  derived a general form of the equation of motion, 
since then usually called Lindblad equation or \emph{Markovian quantum master equation}~\cite{Breuer2007}.  The equation has been successfully applied to many different quantum systems including photons, condensed matter, and quantum computers. The Lindblad equation is still
today
the most popular tool for the investigation of open quantum systems.  Without a doubt, Lindblad was a forerunner of the theory of open quantum systems.

\vspace{2ex}
\noindent\textbf{Quantum stochastic thermodynamics}:
The conventional theory of thermodynamics deals with energetic transaction at the macroscopic scale where the fluctuations of microscopic states are negligible.  However, advances in nanotechnologies have forced us to develop a theory of thermodynamics at the mesoscopic scale where the fluctuations play a significant role. For classical systems, stochastic approaches such as Langevin equations already existed at the time of the Lindblad's book.  However, 
a major advancement in the understanding of the relation between fluctuations and thermodynamics was achieved by the development of \emph{stochastic thermodynamics} near the end of 20th century~\cite{Seifert2012,Sekimoto2010}, wherein
thermodynamic quantities are defined as functionals of a single stochastic trajectory $x(t)$. 

For example,  the work functional is defined by
\begin{equation}
	W[x(\cdot)] = \int_{t_i}^{t_f} \pdv{H(x;\lambda)}{\lambda} \dv{\lambda}{t} dt = \int_{t_i}^{t_f} \pdv{H(x;\lambda(t))}{t} dt
\end{equation}
where $H(x,\lambda)$ is a Hamiltonian with a control parameter $\lambda$~\cite{Sekimoto1997,Jarzynski1997a,Jarzynski2007}.
Thermodynamic
work can be obtained by taking ensemble average over the stochastic trajectories. 

For quantum systems, not only the thermal fluctuation but also the quantum fluctuation can play a role in thermodynamics processes, which is the starting point of
\emph{quantum thermodynamics}~\cite{Alicki2018}. The most common approach is to use the quantum Markovian dynamics obtained from the Lindblad equation in place of classical Markovian dynamics. Lindblad called it \emph{quantum stochastic processes} in Chapter 7. Extending the idea of stochastic thermodynamics to quantum systems,  thermodynamic quantities are considered as a functional of the whole evolution of state $\rho(t)$ obtained from the quantum stochastic processes.  Alicki~\cite{Alicki1979} derived the principles of thermodynamics using this approach. Lindblad followed the Alicki's approach and evaluated the rate of entropy production using the Lindblad equation in Chapter 7. This approach has been extensively used 
since. We can say that Lindblad was a founding father of quantum thermodynamics.

\vspace{2ex}
\noindent\textbf{Fluctuation theorems}:
The most significant outcome of the stochastic thermodynamics is the discovery of fluctuation theorems.  Starting from Bochkov and Kuzovlev~\cite{Bochkov1977}, various forms of fluctuation relations beyond the standard fluctuation-dissipation theorem were discovered~\cite{Sevick-Searles2008,Jarzynski2011,Seifert2012,Gawedzki2013,Spinney2013}.
Among them, the \emph{Jarzynski equality}~\cite{Jarzynski1997a} is 
arguably the most famous and the most relevant to the Lindblad's book.  Suppose that a system in thermal equilibrium at inverse temperature $\beta$ is brought  out of the equilibrium by a time-dependent Hamiltonian.  The corresponding stochastic trajectories start from a point $x_0$ randomly chosen from the Gibbs distribution $\rho(x_0) = e^{-\beta H(x_0)}/Z_i$.  The trajectories are determined by a Hamiltonian or stochastic dynamics. Then, the following equality
between a non-equilibrium average
(left-hand side) and a difference between equilibrium
quantities (right-hand side) holds~\cite{Jarzynski1997a}:
\begin{equation}\label{eq:jarzynski}
	\expval{e^{-\beta W}}_{x_0} = e^{-\beta \Delta F}
\end{equation}
where  $\Delta F = T (\ln Z_f/Z_i)$ is the free energy difference, and the ensemble average $\expval{\cdots}_{x_0}$ is taken over the initial position $x_0$ with the Gibbsian meausure $\rho(x_0)$.  When the system 
goes out of equilibrium,  temperature is no longer defined.  The temperature in the equality is
therefore
the initial temperature of the system.  Using  Jensen's inequality, Eq. \eqref{eq:jarzynski} implies the second law in the form of work vs.free energy inequality:
\begin{equation}\label{eq:W_diss}
	\expval{W} - \Delta F \ge 0.
\end{equation}

Classical fluctuation relations follow from that
the entropy production functional can be 
defined \textit{both} in terms of heat exchanges with
reservoirs, \textit{and} as a log-ratio of
probabilities of forward and time-reversed
processes. Almost tautological relations in the second formulation then yield highly non-trivial equalities (the fluctuation relations) in the first~\cite{Gawedzki2013}. 
An intermediate result is represented by 
Crooks' theorem~\cite{Crooks1999}, 
an exact relation between two Markovian dynamics, one in the forward time and the other in backward time. One consequence is a precise equality between dissipation and irreversibility~\cite{Kawai2007,Parrondo2009}:
\begin{equation}\label{eq:arrowoftime}
	\beta (\expval{W} - \Delta F) =  S(\rho(t)\|\tilde{\rho}(\tau-t))
\end{equation}
where $\rho$ and $\tilde{\rho}$ are the state density of the forward and backward processes, respectively, $\tau$ is the final time of the process, and $t$ can be any time between 0 and $\tau$. This equality is closely related to Eq. \eqref{eq:S(r1|r1')=DS} derived 
by Lindblad in Sec. \ref{sec:synopsis-5}.  Using the same process considered in Sec. \ref{sec:synopsis-5}, we find that $\Delta F=0$,  $t=\tau$. $\rho_1=\rho(\tau)$, $\rho_1^\prime=\tilde{\rho}(0)$ and that Eq. \eqref{eq:arrowoftime} coincides with Eq. \eqref{eq:S(r1|r1')=DS}. 
We hence find that Lindblad was very close to
deriving one type of fluctuation relations.

Jarzynski's equality for closed
quantum systems was to our knowledge
first found by David Thouless in 1996
\cite{Jarzynski-private}, and first presented
in a pre-print of Kurchan in 2000
\cite{Kurchan2000}.
The same derivation goes through if the
open quantum system evolution is unital~\cite{Albash2013,Rastegin2014}. 
The Jarzynski equality for general open
quantum systems encountered various difficulties
extensively discussed in the literature at an early stage~\cite{Campisi2011,Esposito2009}.  
One difficulty is the relation (if any) between
quantum Jarzynski equality and quantum time reversals.
In an environmental
representation such a relation was found by Crooks~\cite{Crooks2008}. In an intrinsic representation
this was found 
in~\cite{Chetrite2012}
for quantum Markov dynamics (Lindblad equation),
but is in general unclear, see
\cite{Aurell2015} for a critical discussion.

\vspace{2ex}
\noindent\textbf{Quantum measurement, Maxwell demon and Landauer principle:}
In Chap. 11,  Lindblad discussed work extraction from a passive state by quantum measurement, seemingly in violation of the second law.   This is precisely a quantum version of the Maxwell's demon.  In modern understanding of Maxwell demon, the decrease of entropy of the system by the action of demon corresponds an increase of information in the demon itself (measurement device). The part of the world considered in the analysis is not brought back to its original state unless this information in the demon is erased.
Lindblad condidered the measurement device (demon) as a macroscopic object, interacting with its own environment at inverse temperature $\beta_0$. A part of energy transferred to the device dissipates as heat $Q$ which corresponds to the erasure of information in the device.  Then, he derived 
\begin{equation}\label{eq:Landauer-like}
	\beta_0 Q = \Delta S_I^M.
\end{equation}
Equation \eqref{eq:Landauer-like} indicates that the information acquired by the device dissipates into its own heat bath.  He pointed out that work extraction by the quantum measurement is essentially the heat engine operating between the two heat baths, one for the system and the other for the device assuming $\beta < \beta_0$.

Landauer~ \cite{Landauer1961} in 1961 proposed that  an erasure of information the demon acquired always satisfies
\begin{equation}
	\label{eq:Landauer-principle}
	\beta Q \ge \Delta S
\end{equation}
where $Q$ is the heat released to a bath and $\Delta S$ is the entropy change of the system. The inequality \eqref{eq:Landauer-principle}, known as \emph{Landauer principle}, suggests that  the heat released into the environment when erasing the record of the demon's actions is enough to match the entropy which the demon manages to decrease~\cite{Bennett1982}. Mandal and Jarzynski~\cite{Mandal2012} discusses a classical model of Maxwell's demon illustrating the mechanism in detail. 

We note that the Lindblad's equality \eqref{eq:Landauer-like} is a special case of the Landauer principle. A process realizing the Landauer principle as an equality must proceed very slowly. For processes completed in a finite time $t$ it was first 
predicted on general grounds \cite{Sekimoto1997}
and found in specific models \cite{Schmiedl2007}, later confirmed
experimentally \cite{Berut2012} and then 
developed further and put in a wider context~\cite{Aurell2011,Aurell2012} that the principle is modified to
\begin{equation}
	\label{eq:revised-Landauer-principle}
	Q \ge k_B T \Delta S + \frac{C}{t}.
\end{equation}
where $C$ in above is a suitable measure of the
distance between the initial and final state.
The formal derivation of quantum Landauer principle is more complicated, see Ref.~\cite{Anders2010}. 

It is not clear if Lindblad was aware of the Landauer principle at the time he wrote the book. He cites
only the much earlier discussion by Szilard.
But we can say that Lindblad's reasoning in Chapter 11 is closely related to the Landauer principle. Furthermore,
from later papers it is clear that Lindblad thought that non-equilibrium thermodynamics and quantum measurement share the similar irreversibility where entanglement spread through the heat bath or macroscopic measurement device~\cite{Lindblad2011}. We therefore 
hold it likely that the similarity of irreversibility between non-equilibrium thermodynamics and quantum measurement later led Lindblad to the Markov chain model of quantum measurement theory.

\vspace{2ex}
\noindent\textbf{P-entropy and entropy production} 
One of the signatures of irreversibility is entropy production. The second law of thermodynamics only states that it is positive for any irreversible process. 
It has been a major topic of modern thermodynamics that
(total) entropy productions is composed of system entropy change ($\Delta S_{sys}$) \textit{and} entropy production in the environment, which in the stochastic thermodynamics literature is usually written $\delta S_{env}$. When the environment is a heat bath (or heat baths) $\delta S_{env}$ is the heat transferred in units of temperature.
Lindblad's central concept of P-entropy (Eq.~\ref{eq:central-eq}) is a clear forerunner of these
ideas, since it can be written as
\begin{equation}
\label{eq:central-eq-version-2}
    S_P[\rho;F]=S_I[\rho]
    + \mathrm{Inf}[\Delta S_{sys} + \delta S_{env};\{\rho_G \in \Omega(F) \}]
\end{equation}
where the infimum is over final Gibbs states
$\rho_G$ which can be reached from $\rho$
by quantum evolutions (work cycles and interactions with baths), using controls in $F$.
$S_I[\rho]$ is the von Neumann (information-theoretic)
entropy in the initial state and 
$S_I[\rho]+\Delta S_{sys}=S_I[\rho_G]$ is the
von Neumann entropy of the final state.

For (classical) over-damped systems the infimum in
Eq.~\ref{eq:central-eq-version-2} is realized as
optimal (classical) transport, a problem also known
as the Monge-Kantorovich problem, or the 
minimum (classical) Wasserstein distance.
This connection between two seemingly very different optimization problems, which leads to the 
revised Landauer principle (Eq.~\ref{eq:revised-Landauer-principle} above), was
apparently first made in the mathematical literature by Otto and Villani~\cite{Otto2000}, and in the physical
literature in \cite{Aurell2011,Aurell2012} as sketched above.
More recent
discussions from a physical perspective can be found in 
\cite{Proesmans2020}, \cite{Nakazato2021}
and \cite{VanVu2022,VanVu2023}.
The extension of Eq.~\ref{eq:central-eq-version-2} to
underdamped Langevin dynamics 
was
carried out in \cite{Muratore2014,Muratore2014b},
and has also been realized in experiments~\cite{Dago2021,Dago2022}.
Optimality for finite-time processes with discrete states
was discussed in
\cite{Muratore2013}, \cite{Remlein2021,Dechant2022}
and \cite{VanVu2022,VanVu2023}
where the latter two also consider generalizations to
quantum processes. 
The quantum Wasserstein (transport) distance as such has
been extensively discussed 
in the information-theoretical
\cite{DePalma2021},
physical~\cite{Friedland2022,Bistron2023}
and mathematical 
\cite{Agredo2017,Cagliotti2021,Gao2021}
literature.

Still, it is obvious that the full extent of 
Lindblad's proposal embiodied by
Eq.~\ref{eq:central-eq-version-2} 
has not been explored. 
It is not clear if also in the
quantum case this quantity has the alternative interpretation as
a geometric (transport) distance between states, and the study of the interplay between optimality and a
limited control set indicated by $F$ is certainly only in its infancy.  We believe 
constructing theories along these lines 
important future tasks for non-equilibrium quantum
thermodynamics.

Another, perhaps equally important direction, can be found in Chap. 5 where Lindblad concluded that the dissipation of heat during reversible processes are due to the formation of global correlations.  A modern theory of entropy production, that is valid even far-from equilibrium, indicates that the formation of correlation inside the heat bath is the main source of the entropy production~\cite{Ptaszynski2019}.  Once the entanglement spreads, it is difficult to determine the phase information by measurement and consequently it is difficult to construct a time-reversed operation, which Lindblad claimed to be the origin of the irreversibility. 
Lindblad's ideas on entropy production were hence consistent with the modern notions.  Furthermore, 
Lindblad thought that the irreversibility of quantum measurement can be explained by the same reasoning.
These both seem to us very promising future directions.

\vspace{2ex}
\noindent\textbf{Measurement-driven engines:}
As Lindblad suggested, quantum measurement can be thought of a heat engine.   In another sense, a measurement device behaves like a heat bath~\cite{Jordan2020}, suggesting that one of heat baths used in conventional thermal engine can be replaced with a quantum measurement. 
More than thirty years after the publication of Lindblad's book, measurement-driven engines extracting work out of a single passive state,
even without feedback control, became a hot research topic~\cite{Elouard2017,Elouard2018,Yi2017}.  These engines are seemingly violating the second law unless the erasure of information is explicitly taken into account, i.e., the quantum version of the Landauer principle, as discussed above. 
It has been also shown that the strong local passivity~\cite{Frey2014} (aka. CP-passivity~\cite{Alhambra2019}) can be broken by the so-called  \emph{quantum energy teleportation} (QET) protocol consisting of a local quantum measurement followed by local operations and classical communication (LOCC)\cite{Alhambra2019,Frey2014}. QET was recently confirmed in experiments~\cite{RodriguezBriones2023}. 
It is a regrettable fact that Lindblad's theory on measurement-driven work extraction has been overlooked 
in most recent papers on the topic, perhaps because his work was too far ahead of his time. We hope that this perspective will help rectify this state of affairs.

\vspace{2ex}
\noindent\textbf{Entanglement-driven engines:}
In Chapter 5 Lindblad shows that the global correlation in the heat bath induced by work is responsible for the entropy increase. It has been shown that the quantum correlation (entanglement or discord) affects thermodynamics significantly. It is now well-known that the quantum correlation can be used as thermodynamic resources. If a system carries quantum correlation, it can be used to drive thermal machines consuming the mutual information stored in the correlation as fuel. In other words, the mutual information can be used in place of work, see for example Ref.~\cite{Bera2017}.  Among many proposed coherence-driven machines, one seemingly violating the second law is spontaneous heat flow from a cold to a hot reservoir under the presence of quantum correlation between the reservoirs~\cite{Partvi2008}.  Based on it, a heat pump entirely driven by quantum entanglement was proposed~\cite{Holdsworth2022}.  The second law is recovered if the mutual information $I$ is assumed to be equivalent to $\beta W$ in the regular heat pump. 
In summary, 
advanced concepts of quantum information are important components of quantum thermodynamics,
that is of thermodynamics extended to the microscale and 
to the quantum regime.

\section{Concluding remarks}
\label{sec:concluding-remarks}

Lindblad introduced unique ideas 
to advance the understanding of non-equilibrium states.
They are grounded in the experimental situation where the experimenter has limited options. He also developed many pieces of new concepts necessary to derive an entropy function for such states. His final proposal, the P-entropy, is closely related to concepts which have
only recently come to the fore in quantum theory.
Unfortunately, neither Lindblad's definition of entropy nor other ideas was noticed at the time of publication.  Since then, research on the non-equilibrium thermodynamics has made great progress. Nevertheless, 
it turns out that many of Lindblad's ideas were reinvented in the modern development of quantum thermodynamics, 
though
his work was almost completely overlooked.  We think that Lindblad was a deep thinker and truly ahead of his time.  We hope that our extended synopsis and modern perspective helps established researchers and young researchers alike to discover his book.

\section*{Acknowledgments}
We thank numerous colleagues for constructive remarks on earlier versions of this text, in particular 
Gavin Crooks,  
Christopher Jarzynski, David Lacoste,
Paolo Muratore-Ginanneschi,
Keiji Saito, Udo Seifert, Ken Sekimoto, Emmanuel Trizac
and Mark Wilde.
The responsibility for remaining omissions and/or
inaccuracies are ours only.
EA acknowledges support of the Swedish Research Council through grant 2020-04980.

\end{document}